\documentclass[12pt,preprint2]{aastex}

\slugcomment{To appear in ApJ Letters}

\shorttitle{Shock chemistry in the outflow of M82}
\shortauthors{Garc\'{\i}a-Burillo et al.}

\begin{document}


\title{SiO chimneys and supershells in M82
\footnote{Based on observations carried out with the IRAM Plateau de Bure 
Interferometer. IRAM is supported by INSU/CNRS (France), MPG (Germany) and IGN (Spain).}}


\author{S.Garc\'{\i}a-Burillo \\
	J.Mart\'{\i}n-Pintado \\
	A.Fuente}

\affil{Observatorio Astron\'omico Nacional (OAN), Campus Universitario \\ Apdo.1143, 
Alcal\'a de Henares, E-28800, Madrid, SPAIN}

\email{burillo@oan.es, martin@oan.es, fuente@oan.es}

%\author{C. D. Biemesderfer\altaffilmark{4,5}}
%\affil{National Optical Astronomy Observatories, Tucson, AZ 85719}
%\email{aastex-help@aas.org}

\and

\author{R. Neri}
\affil{Institut de Radio Astronomie Millim\'etrique (IRAM) \\
300, Rue de la Piscine, 38406-St.Mt.d'H\`eres, FRANCE}
\email{neri@iram.fr}

\begin{abstract}

In this letter we present the first images of the emission of SiO and H$^{13}$CO$^{+}$  
in the nucleus of the starburst galaxy M\,82. 
Contrary to other molecular species, which mainly trace the distribution of 
the star forming molecular gas within the disk, the SiO emission 
extends noticeably out of the galaxy plane. The bulk of the SiO emission is restricted to 
two major features. The first feature, referred to as the SiO supershell,
is an open shell of 150\,pc diameter, located 120\,pc west from the galaxy center.
The SiO supershell represents the inner front of a molecular shell expanding at
$\sim$40\,kms$^{-1}$, produced by mass ejection around a supercluster of young stars containing 
SNR 41.95+57.5. The second feature is a vertical filament, referred to as the SiO 
chimney, emanating from the disk at 200\,pc east from the galaxy center. 
The SiO chimney reaches a 500\,pc vertical height and it is associated with the most 
prominent chimney identified in radiocontinuum maps. 
The kinematics, morphology, and fractional abundances of the SiO gas features in M\,82  
can be explained in the framework of shocked chemistry 
driven by local episodes of gas ejection from the starburst disk. The SiO emission 
stands out as a privileged tracer of the disk-halo interface in M\,82. 
We speculate that the chimney and the supershell, each injecting  $\sim$10$^{7}$M$_{\sun}$ of
molecular gas, are two different evolutionary stages in the outflow phenomenon 
building up the gaseous halo.

\end{abstract}


\section{Introduction}
The detection of large-scale outflows in over a dozen starburst 
galaxies has confirmed the overall predictions of the galactic wind model, first 
proposed by \citet{che85}. It is widely accepted that the driving mechanism of the 
outflow phenomenon in starbursts is linked to the creation of expanding shells of hot gas 
by supernovae. At the end of the starburst cycle, the resulting high supernova rate 
creates a rarefied wind of hot gas in the disk (with temperatures of $\sim$10$^7$K) at several 
thousands of kms$^{-1}$. 
Different {\it hot bubbles} could merge and blow out into the halo, entraining surrounding 
cold gas and dust at several hundreds of  kms$^{-1}$. These bubbles, although initially spherical, 
may evolve eventually into vertical chimneys of gas (Norman \& Ikeuchi 1989; Koo \& McKee 1992; Alton, 
Davies, \& Bianchi 1999). 
The halo outflow would be built up in the end by the local injection of gas from the disk. 
The details of this secular process, which should drive large-scale shocks in 
the molecular gas, remain unknown however, partly because we lack of observational constraints. 

M\,82 is the closest galaxy experiencing a massive star forming episode \citep{rie80, wil99}. 
Its nuclear starburst, located in the central 1\,Kpc, has been studied in virtually all 
wavebands from X-rays to the radio domain. 
X-rays and optical observations have shown the existence of a large-scale biconical outflow of hot gas 
coming out of the plane from the nucleus of M82 (Bregman, Schulman, \& Tomisaka 1995; Shopbell \&
Bland-Hawthorn 1998). This massive outflow is also 
observed at large scales in the cold gas and dust \citep{sea01,alt99}.
At smaller scales, close to the disk-halo interface of M\,82, there is observational 
evidence of local sources of gas injection. Using 1.4 and 5 GHz VLA data, \citet{wil99} have shown 
the existence of a large (diameter $\sim$120 pc) expanding shell of ionized gas, close to the 
supernova remnant SNR 41.95+57.5. A molecular gas counterpart of this supershell was tentatively identified 
by \citet{nei98} and later discussed in \citet{wei99,wei01} and \citet{mat00}. \citet{wil99} 
detected also the signature of four chimneys of hot gas. The most 
prominent one, located on the northeastern side, reaches a vertical height of $\sim$200\,pc. The 
molecular gas counterpart of the chimney has remained so far undetected (see \citet{wei01}).     

In this letter we present the first high-resolution ($\sim$5'') image of the emission
of silicon monoxide (SiO) in the nucleus of M\,82. 
SiO is known to be a privileged tracer of large-scale shocks in the interstellar medium of galaxies 
\citep{mpi97, gbu00, gbm01}. The SiO emission is used in this work to study the occurrence of 
shocks in the halo-disk interface of M\,82. We present the first evidence of a SiO expanding 
supershell, related with the superbubble of hot gas, and 
the first detection of a molecular gas chimney associated with a violent mass ejection event in 
M\,82.

\section{Observations}

Observations of M\,82 were completed with the IRAM array at
Plateau de Bure (France) during 1999 June. We observed simultaneously
the J=2--1 line of SiO (86.847\,GHz), and the J=1--0 line of 
H$^{13}$CO$^+$ (86.754\,GHz), using the CD set of configurations. The 55$''$ primary beam 
field of the array was phase-centered at $\alpha_{J2000}$=$09^h55^m51.9^s$ and 
$\delta_{J2000}$=$69^{\circ}40'47.1''$, which corresponds to the 2.2$\mu$m peak (Joy, Lester, \& Harvey 1987). 
We adjusted the spectral correlator to give a contiguous bandwidth of 1500\,kms$^{-1}$. 
The frequency resolution was set to 2.5\,MHz (8.64\,kms$^{-1}$). We calibrated visibilities 
using as amplitude and phase references 0836+710 and 0716+714. The absolute flux 
scale and receiver passband shape were derived on MWC\,349 
and 3C\,273, respectively. Cleaned maps are $256\times 256$ pixels in extent, with a pixel 
size of $0.6''$. The synthesized beam is almost circular ($5.9''\times 5.6''$, PA=105$^{\circ}$). 
The rms noise level in 2.5\,MHz wide channel maps, derived after subtraction of the 
continuum emission, is 1\,mJy/beam (5\,mK). We take as distance to M\,82, D=3.9\,Mpc \citep{sak99}; 
the latter implies 1''$\sim$20\,pc.

\section{Results}

Fig.1 shows the velocity-integrated intensity map of SiO(2-1) in the central 
1\,Kpc of M\,82. Two specific regions in the nucleus of M\,82 contain the bulk of the SiO 
emission, which extends noticeably out of the galaxy plane. On the eastern side of the disk, 
a vertical feature emanates from the disk at $(\Delta\alpha,\,\Delta\delta)\sim$(+10'',\,+5''), 
indicating the blow-out of molecular gas out of the plane of M\,82. This feature extends 
out to the edge of the primary beam on both sides of the plane along PA$\sim$--40$^{\circ}$, 
reaching a vertical height of $\sim$500\,pc northwards. It is however unresolved 
transversely. The second dominant feature in the map, centered at (-5'',\,-2''), is an arc-like structure 
closed on its southern hemisphere. Two emission peaks, $\sim$150\,pc apart, define
the arc diameter measured along the major axis. We describe below these two prominent 
sources of the SiO map, hereafter referred to as the {\it chimney} and the {\it supershell},
and discuss their relation with the local mass ejection signatures observed in the disk of M\,82.

\subsection{The SiO chimney}

The SiO chimney is closely related to a radiocontinuum (RC) chimney identified by \citet{wil99}, 
which represents a signature of ejection of ionized gas. The RC chimney in Fig.1 is delimited by two black-shaded 
emission blobs on the northern edge of the continuum map. The filament stands out due to free-free 
absorption by the ionized gas lying inside. Although the RC chimney is not the only ejection feature of ionized 
gas in M\,82, it is the most significant blow-out signature. Our detection of SiO along this chimney shows that 
molecular gas can be entrained and survive up to high vertical z-distances ($\sim$500\,pc). 
The $^{12}$CO(2--1) interferometer map of \citet{wei01} does not show a clear signature linked to the SiO 
chimney. Although there is an emission peak in the eastern end of the map, near the southern end 
of the SiO chimney, \citet{wei01} have discarded it as an artifact. At much larger scales 
however ($\sim\pm$2-3\,Kpc), the low-resolution $^{12}$CO maps of the galaxy's halo \citep{thu00,sea01} show
evidence that a giant molecular outflow exists in M\,82.
Contrary to the high-density gas (n(H$_2$)$>$10$^{4-5}$cm$^{-3}$) probed locally 
by the SiO emission along the chimney, the halo gas traced by $^{12}$CO at larger scales is mostly 
diffuse (n(H$_2$)$\sim$10$^{3}$cm$^{-3}$, see \citet{sea01}). This is expected if the SiO emission 
arises from a thin layer of shocked gas (see below).

The SiO radial velocities measured along the chimney indicate that the gas flow is driven by 
an ejection event, rather than by the rotation of the galaxy. 
Fig.2 shows the channel maps for three velocity intervals containing the bulk of the SiO 
emission. The chimney feature stands out in the 100-240\,kms$^{-1}$ range. The strongest 
SiO emission on the northern and southern segments of the chimney appears at v$\sim150\,$kms$^{-1}$, 
i.e. at {\it forbidden} velocities, $\sim$100\,kms$^{-1}$ blueshifted relative to the systemic velocity 
(v$_{sys}\sim$225$\pm$20\,kms$^{-1}$). The emission from a filament of ionized gas lying near the SiO chimney 
has been studied by \citet{sbh98}, using H$\alpha$+NII lines. Most interestingly, the bluest 
component detected in the optical lines appears also at v$\sim150\,$kms$^{-1}$.
The coincidence in radial velocities found between SiO and H$\alpha$+NII
suggests that the working surfaces for molecular and ionized gas share a similar 
geometry. 
 
The SiO chimney is also detected in H$^{13}$CO$^{+}$ at similar velocities, 
although at a significantly lower level (see Fig.2). The H$^{13}$CO$^+$ emission is strongest in the 
southern segment of the chimney, at both {\it forbidden} and {\it permitted} velocities.  
H$^{13}$CO$^+$ is best detected at the base of the chimney, close to the galaxy disk. 
We will adopt the same approach to the one used by \citet{gbu00} in NGC\,253 to 
estimate the molecular mass of dense gas (M(H$_2$)) and the SiO fractional abundances (X(SiO)) 
in the chimney. Assuming that the physical parameters of the SiO gas are similar 
in M\,82 and in NGC\,253 (n(H$_2$)=1-10$\times$10$^{5}$cm$^{-3}$ and T$_k$=50K), 
we derive the SiO/H$^{13}$CO$^+$ abundance ratio that fits the observed SiO/H$^{13}$CO$^+$ 
line ratio, using a Large Velocity Gradient (LVG) model. If we 
adopt a canonical value for X(H$^{13}$CO$^+$)=10$^{-10}$, the implied SiO abundance would range from 
$\sim$2\,10$^{-10}$ to $>$3.5\,10$^{-10}$.  The SiO abundance is significantly 
enhanced in the chimney, surpassing by $\sim$1--2 orders of magnitude the typical value of quiescent 
gas environments (Ziurys, Friberg, \& Irvine 1989). This local value of the SiO abundance is also 
remarkably larger than the global estimate of X(SiO) for the whole M\,82 disk ($\sim$10$^{-11}$), 
first derived by \citet{sag95}. Conclusions on the enhancement of the SiO 
abundance in the chimney are strengthened if we assume a lower density for the gas: 
for n(H$_2$)=5$\times$10$^{4}$cm$^{-3}$, we derive X(SiO)$\sim$10$^{-9}$ (see detailed discussion 
in \citet{gbu00}). The present estimates indicate that shock chemistry 
is heavily processing molecular gas upon ejection. We can derive the M(H$_2$) mass contained in the chimney 
by integrating the LVG-based estimate of N(H$_2$). We obtain M(H$_2$)$\sim$6\,10$^{6}$M$_{\sun}$. 

\subsection{The SiO supershell}

The SiO supershell is associated with similar signatures of gas ejection located 
around SNR 41.95+57.5. \citet{wil99} have reported the existence of an expanding shell of 
ionized gas of $\sim$120\,pc diameter, near SNR 41.95+57.5, which is also identified 
in the NeII map of \citet{ach95}. \citet{wei99, wei01} and \citet{mat00} found in their 
$^{12}$CO interferometer maps further evidence of an expanding supershell of molecular gas around 
SNR 41.95+57.5. The maps of the ionized and molecular supershells show morphological differences, 
however. The emission of ionized gas seems to lie inside the CO supershell. 
Like the ionized gas, the SiO supershell (of $\sim$150\,pc diameter) seems also smaller than 
its CO counterpart (of $\sim$200\,pc diameter). The southern wall of the SiO supershell 
represents the inner front of the CO shell which is also seen to protrude southwards in the map 
of \citet{wei01}. On larger scales, \citet{wil99} have identified a spur-like 
filament extending into the north from the supershell center in the disk. The morphology of the 
SiO supershell, closed southwards, confirms that the ejection has broken the primary molecular 
shell to the north. 

The kinematics of SiO in the supershell are roughly consistent with the expansion 
scenario depicted by \citet{wei99, wei01}. The SiO emission at 
the center of the supershell (at $\sim$(-5'',\,-2'')) show hints of a
double-peaked profile. The strongest (weakest) component at v$\sim$100\,kms$^{-1}$ (v$\sim$180\,kms$^{-1}$) 
would represent the approaching (receding) side of a molecular shell in the galactic disk, 
expanding at v$_{exp}\sim$40\,kms$^{-1}$. As expected for an expanding shell, 
the SiO spectrum at its southern end (at $\sim$(-2'',\,-8'')) has one 
component centered at v$\sim$140\,kms$^{-1}$. 

%Further evidence of double-peaked profiles in the SiO 
%spectra is found at the northern end of the shell (at $(\Delta\alpha,\,\Delta\delta)\sim$(-12'',\,3'')). 

Under the same assumptions of section 3.1 we can estimate both the mass of dense molecular gas 
M(H$_2$) and the fractional abundance of SiO in the supershell. 
The inferred SiO abundances are X(SiO)$\sim$0.4--1\,10$^{-10}$, namely a factor of $\sim$4 lower 
than in the SiO chimney. Processing of molecular gas by shocks seems milder in the supershell, 
especially within the galaxy disk where the abundance of SiO approaches values typically found in 
Photon-Dominated Regions (PDR) (X(SiO)$\sim$10$^{-11}$; see Walmsley, Pineau des For\^ets, \& Flower (1999)). 
The derived mass of molecular gas in the SiO supershell is M(H$_2$)$\sim$1.6\,10$^{7}$M$_{\sun}$, 
roughly in agreement with the value derived by \citet{wei99}, using $^{12}$CO(2--1) data.

\section{Discussion and Conclusions}

Numerical simulations studying the evolution of outflows in starbursts 
\citep{tom93, suc96} have predicted that the resulting hot galactic wind, observed in X-rays and 
diffuse H$\alpha$ emission, may drag the cooler and denser material of the blown-out disk up 
to 1--2 kpc above the plane of a galaxy. These models show that, near the base of the outflow, 
at a scale height of $\sim$500\,pc, filaments of cold disk material should be present. The detection 
of SiO emission from a prominent chimney and a giant shell in M\,82 provides a nice confirmation 
of these models, proving that the entrained cold material can survive in molecular form in spite 
of the high-velocities (several hundreds kms$^{-1}$ in the SiO chimney) involved upon ejection. 
The chemical processing of dust grains by shocks can be at work 
during the blow-out of the disk; the latter naturally explains the high abundances of SiO 
measured in the molecular gas of the chimney, and to a lesser extent, in the supershell. 
Contrary to other species, which mainly trace the distribution of the star forming molecular 
gas within the disk, SiO stands out as a privileged tracer of the disk-halo interconnection 
in M\,82.

Several scenarios can be envisaged to account for the different morphologies and properties 
characterizing the chimney and the supershell. We discuss the case where the differences 
might reflect the evolution expected for an ejection event from the disk.  
Furthermore, the energies required to form the chimney or the supershell may largely differ. 
Their unlike morphologies may also result from the action of collimating 
mechanisms shaping differently the outflow of molecular gas.

In the frame of the evolutionary scenario, the SiO chimney, extended up to 500\,pc above the galaxy 
plane, would be an evolved ejection episode. In contrast, the supershell (of $\sim$75\,pc radius) 
would be just starting to undergo a blow-out. \citet{wil99} have come to similar conclusions 
analyzing the morphologies of the related radiocontinuum features. Additional insight is gained
by estimating the kinematical ages of the SiO structures. The reported size 
($\sim$75\,pc radius) and expansion velocity ($\sim$40$\pm$5\,kms$^{-1}$) allows to infer an age of 
$\sim$2\,10$^{6}$ years for the SiO supershell. This is probably an upper limit, as the gas 
flow has been likely decelerated during the expansion, as already pointed out by \citet{wei99}.
Deriving the kinematical age of the chimney is less straightforward, however.    
Although the radial velocities measured along the chimney reveal an ejection-dominated flow, 
the value of deprojected velocities are strongly model-dependent. A global model for the molecular 
gas outflow needs a complete high-resolution mapping of the molecular halo in M\,82. 
However, we found in our data evidence that the dynamics of the entrained molecular 
gas is similar to the ionized gas near the location of the SiO chimney. If the model of \citet{sbh98} 
(based on a global fit on several optical filaments) held for the SiO chimney, the estimated 
ejection velocity would be $\sim$500\,kms$^{-1}$. The derived kinematical age for the SiO chimney 
($\sim$10$^{6}$ years) would not be significantly different from the one determined for the 
supershell. As the evolutionary link hypothesis depicted above is probably correct, this result
suggests that the time-scale for evolution is shorter in the chimney than in the supershell. 
This may indicate that the energy required to create the chimney is comparatively larger.

We estimate that the kinetic energy contained in the SiO supershell is E$_{kin}\sim$2\,10$^{53}$ergs. 
Assuming that the typical type-II supernova energy input is 10$^{51}$ergs, and that 
only $\sim$10$\%$ of the explosion energy is transferred  to kinetic energy (see numerical models of 
\citet{che74}), the formation of the supershell would require $\sim$2\,10$^{3}$ correlated explosions in 10$^{6}$ years. 
Most notably, the derived kinetic energy of the SiO chimney is much larger: 
E$_{kin}\sim$10$^{55}$ergs (taking $\sim$500\,kms$^{-1}$ as ejection velocity); this would  
require 10$^{5}$ correlated supernovae in 10$^{6}$ years, namely a supernova rate of 
0.1SN yr$^{-1}$. This result holds even in the improbable scenario where the outflow is nearly 
parallel to the galaxy plane. In this case we obtain a lower limit of E$_{kin}\sim$4\,10$^{53}$ergs 
for the chimney. This is still a factor 2-3 larger than derived for the supershell. 
Although the energy requirements to form the SiO chimney might be very stringent if gas velocities 
are close to hundreds of kms$^{-1}$, the measured supernova rate in the central disk of 
M\,82 (0.1SN yr$^{-1}$; see Kronberg, Biermann, \& Schwab (1985)) may account for it.     

We can speculate that the large-scale molecular gas halo of M\,82 detected by \citet{sea01}, 
has been built up by local episodes of gas injection from the disk. The two SiO features
reported in this letter, injecting each $\sim$10$^{7}$M$_{\sun}$ gas, 
provides a tantalizing evidence that this secular process is at work in M\,82. 
\citet{sea01} estimated a molecular gas mass of the halo of $\sim$5\,10$^{8}$M$_{\sun}$; the latter
implies we would need 20-50 of these local episodes to build up the halo. In view of 
the available estimates for the age of the starburst (5\,10$^{7}$-10$^{8}$ years), and the energy  
deposited by supernovae explosions during this time ($\sim$10$^{58}$ ergs) we can conclude that 
both the time-scales and the energy input required to form 20-50 of these ejection episodes 
are well accounted for by the starburst engine in M\,82.

\acknowledgments

This work has been partially supported by the Spanish DGES under grant number AYA2000-927
and CICYT-PNIE under grant PNE014-2000-C. We heartily thank Karen Wills and Axel Weiss for
providing free access to their data. We acknowledge the IRAM staff from 
the Plateau de Bure and from Grenoble for carrying the observations and help provided 
during the data reduction.
 



\clearpage

\figcaption{The velocity-integrated intensity map of SiO(v=0,J=2-1), in the 
central region of M\,82(levels: 0.10 to 0.35\,Jy\,beam$^{-1}$kms$^{-1}$ by steps of 
0.05\,Jy\,beam$^{-1}$kms$^{-1}$; 1$\sigma$=0.040\,Jy\,beam$^{-1}$kms$^{-1}$), is overlaid with the radiocontinuum emission image at 
4.8GHz from \citet{wil99} (gray-scale saturated from 1.5E-04\,Jy\,beam$^{-1}$ to 4E-04\,Jy\,beam$^{-1}$).
The outer circle delimits the Bure primary beam field at 87 GHz (55''). 
The synthesized beam ($5.9''\times5.6''$) is pictured on the bottom left corner. 
A white line traces the galaxy major axis at PA$_{disk}$=70$^{\circ}$. 
$(\Delta\alpha,\,\Delta\delta)$ offsets are referred to the phase tracking center. 
The starred and filled squared markers show the positions of the dynamical center and the 
SNR41.95+57.5, respectively. The location of the radiocontinuum filament (RC) is highlighted 
by an arrow. 
\label{fig1}}

\figcaption{We compare the emission of SiO(v=0,J=2-1) (contour levels: -0.09, 0.09 to 
0.35\,Jy\,beam$^{-1}$kms$^{-1}$ by steps of 0.025\,Jy\,beam$^{-1}$kms$^{-1}$) and H$^{13}$CO$^+$ 
(gray scale: 0.09, 0.12 to 0.90\,Jy\,beam$^{-1}$kms$^{-1}$ by steps of 0.055\,Jy\,beam$^{-1}$kms$^{-1}$) 
integrated in three adjacent velocity intervals (shown in the top right corner), towards the center 
of M\,82. The major axis and dynamical center are identified as done in Fig. 1.\label{fig2}}

\clearpage 

\plotone{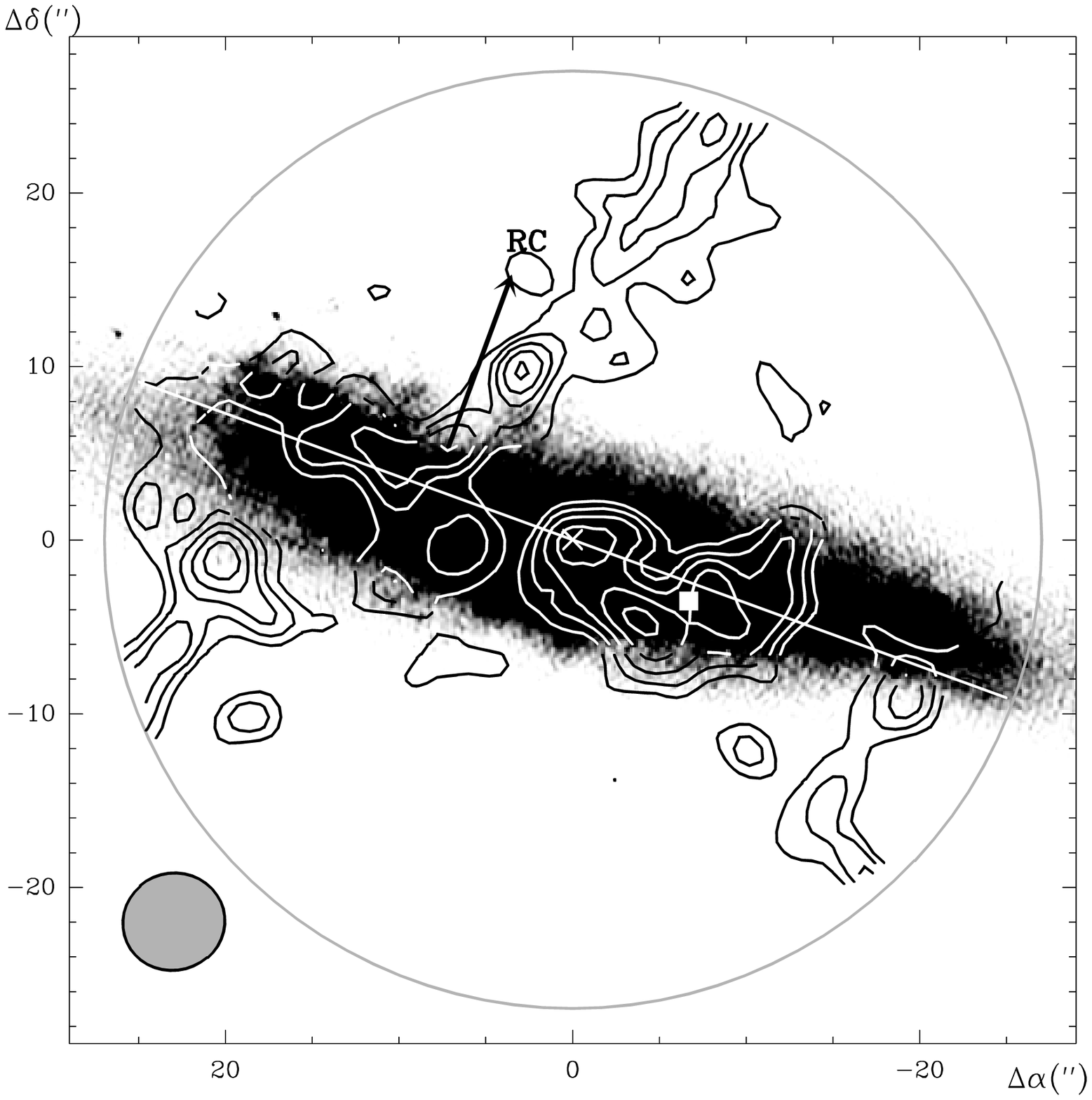}

\plotone{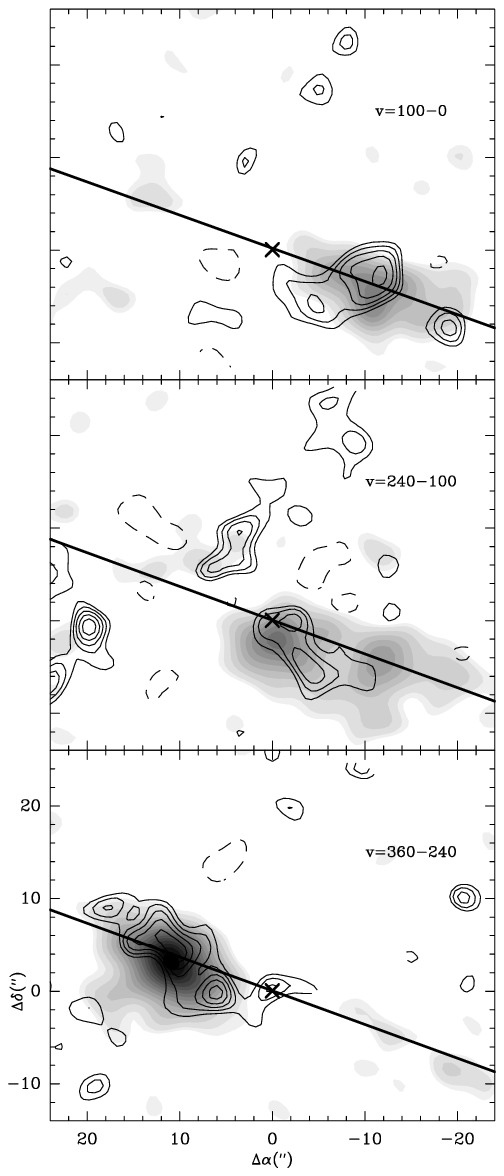}

%% If you are not including electonic art with your submission, you may
%% mark up your captions using the \figcaption command. See the 
%% No more than seven \figcaption commands are allowed per page, 
%% so if you have more than seven captions, insert a \clearpage 

%% Tables should be submitted one per page, so put a \clearpage before

%% Two options are available to the author for producing tables:  the
%% deluxetable environment provided by the AASTeX package or the LaTeX
%% table environment.  Use of deluxetable is preferred.
%%

%% Three table samples follow, two marked up in the deluxetable environment,
%% one marked up as a LaTeX table.

%% In this first example, note that the \tabletypesize{}
%% command has been used to reduce the font size of the table.
%% Note also that the \label command needs to be placed 

%% The following command ends your manuscript. LaTeX will ignore any text


\begin{thebibliography}{}

\bibitem[Achtermann \& Lacy(1995)]{ach95} Achtermann, J. M., \& Lacy, J. H. 1995, \apj, 439, 163

\bibitem[Alton et al.(1999)]{alt99} Alton, P. B., Davies, J. I., \& 
Bianchi, S. 1999, A\&A, 343, 51

\bibitem[Bregman et al.(1995)]{brg95} Bregman, J. N., Schulman, E., \& 
Tomisaka, K. 1995, \apj, 439, 155

\bibitem[Chevalier (1974)]{che74} Chevalier R. A. 1974, ApJ, 188, 501

\bibitem[Chevalier \& Clegg (1985)]{che85} Chevalier R. A., \& Clegg, A. W. 1985, 
Nature, 317, 44   

\bibitem[Garc\'{\i}a-Burillo et al.(2000)]{gbu00} Garc\'{\i}a-Burillo, S., 
Mart\'{\i}n-Pintado, J., Fuente, A., \& Neri, R. 2000, A\&A, 355, 499

\bibitem[Garc\'{\i}a-Burillo \& Mart\'{\i}n-Pintado(2001)]{gbm01} Garc\'{\i}a-Burillo, S., \&  
Mart\'{\i}n-Pintado, J. 2001, The Promise of FIRST, ESA SP-460, ed. by Pillbratt G. et al., in press


\bibitem[Joy et al.(1987)]{joy87} Joy, M., Lester, D. F., \& Harvey, P. M. 1987, \apj, 319, 314 

\bibitem[Koo \& McKee(1992)]{kmc92} Koo, B. C., \& McKee, C. F. 1992, \apj,
388, 93

\bibitem[Kronberg et al.(1985)]{kro85} Kronberg, P. P., Biermann, P., \& Schwab, F. R. 1985, \apj,
291, 693

\bibitem[Mart\'{\i}n-Pintado et al.(1997)]{mpi97} Mart\'{\i}n-Pintado, J., de Vicente, P., 
Fuente, A., \& Planesas, P. 1997, \apjl, 482, L45

\bibitem[Matsushita et al.(2000)]{mat00} Matsushita, S., Kawabe, R., Matsumoto, H., Tsuru, T. G., 
Kohno, K., Morita, K., Okumura, S. K., \& Vila-Vilar\'o, B. 2000, \apjl,
545, L107

\bibitem[Neininger et al.(1998)]{nei98} Neininger, N., Gu\'elin, M., Klein, U., 
Garc\'{\i}a-Burillo, \& Wielebinski, R. 1998, A\&A, 339, 737
   

\bibitem[Norman \& Ikeuchi(1989)]{nik89} Norman, C. A., \& Ikeuchi, S. 1989, \apj,
345, 372

\bibitem[Rieke et al.(1980)]{rie80} Rieke, G. H., Lebofsky, M. J., Thompson, R. I., 
Low, F. J., \& Tokunaga, A. T. 1980, \apj, 238, 24

\bibitem[Sage \& Ziurys(1995)]{sag95} Sage, L. J., \& Ziurys, L. M. 1995, 
\apj, 447, 625

\bibitem[Sakai \& Madore(1999)]{sak99} Sakai, S., \& Madore, B. F. 1999, 
\apj, 526, 599

\bibitem[Seaquist \& Clark(2001)]{sea01} Seaquist, E. R., \& Clark, J. 2001, 
\apj, 552, 133

\bibitem[Shopbell \& Bland-Hawthorn(1998)]{sbh98} Shopbell, P. L., \& 
Bland-Hawthorn, J. 1998, \apj, 493, 129

\bibitem[Suchkov et al.(1996)]{suc96} Suchkov, A. A., Berman, V. G., Heckman, T. M., \& 
Balsara, D. S. 1996, \apj, 463, 528

\bibitem[Tomisaka \& Bregman(1993)]{tom93} Tomisaka, K., \& Bregman, J, N. 1993, \pasj, 45, 513  


\bibitem[Thuma et al.(2000)]{thu00} Thuma, G., Neininger, N., Klein, U., \& Wielebinski, R. 2000, A\&A, 
358, 65

\bibitem[Walmsley et al.(1999)]{wal99} Walmsley, C. M., Pineau des For\^ets, G., \& Flower, D. R. 
1999, A\&A, 342, 542 

\bibitem[Weiss et al.(1999)]{wei99} Weiss, A., Walter, F., Neininger, N., \& 
Klein, U. 1999, A\&A, 345, L23

\bibitem[Weiss et al.(2001)]{wei01} Weiss, A., Neininger, N., H\"uttemeister, S., 
\& Klein, U. 2001, A\&A, 365, 571
  

\bibitem[Wills et al.(1999)]{wil99} Wills, K., Redman, M. P., Muxlow, W. B., \& 
Pedlar, A. 1999, \mnras, 309, 395

%\bibitem[Yun et al.(1993)]{yun93} Yun, M. S., Ho, P. T. P., \& Lo, K. Y. 1993, \apj, 411, L17

%\bibitem[Yun et al.(1994)]{yun94} Yun, M. S., Ho, P. T. P., \& Lo, K. Y. 1994, Nature, 372, 530

\bibitem[Ziurys et al.(1989)]{ziu89} Ziurys, L. M., Friberg, P., \& Irvine, W. M. 1989, 
\apj, 343, 201

\end{thebibliography}
\end{document}